\documentclass{kluwer}    
\usepackage[dvips]{graphicx}

\newdisplay{guess}{Conjecture}

\begin{document}                                                                                   
\begin{article}
\begin{opening}         
\title{X-ray Periodicity in AGN} 
\author{Karen M.\ \surname{Leighly}}  
\runningauthor{Karen M.\ Leighly}
\runningtitle{X-ray Periodicity in AGN}
\institute{The University of Oklahoma}

\begin{abstract}
Significant (marginal) detections of periodic signals have been
recently reported in 3 (4) Active Galactic Nuclei.  Three of the
detections were obtained from long {\it EUVE} light curves of
moderate-luminosity Seyfert galaxies; the fourth was discovered in
{\it Chandra} data from the low-luminosity Seyfert 1 galaxy NGC~4395.
When compared with Cyg~X-1, I find that the period is related to the
luminosity as $P\propto L^{2/3}$ rather than the expected one-to-one
relationship. This result might be explained if the QPO is associated
with the inner edge of the optically thick accretion
disk, and the inner edge radius depends on the source luminosity (or
black hole mass).  A discussion of uncertainties in the period
detection methodology is also discussed.

\end{abstract}

\end{opening}           

\section{Introduction}  

Quasiperiodic oscillations are a characteristic feature of solar-mass
black hole X-ray variability, especially when  the objects are in the
very high (or transition) state.  Active Galactic Nuclei are similar
to stellar-mass black holes in that they are both powered by accretion
onto a central black hole.  Therefore, it seems plausible, or at least
possible, that quasiperiodic oscillations should be observed in active
galaxies as well. 

The search for periodic signals in AGN light curves has a long and
checkered history, however.  A number of claims of periodicity have
been put forward, only to later be proved faulty.  The problems fall
into three categories.  The first and most common problem involves
claims in which the confidence level of the detection is
overestimated.   An example is the recent claim of a periodic signal
with frequency of $2.4 \times 10^{-4}\rm \, Hz$ in {\it XMM-Newton}
data from Mrk~766 (Boller et al.\ 2001).  It was later shown by
Benlloch et al.\ (2001) that the simulated light curves used to establish the
significance of the detection were not properly randomized.  The
second problem involves detector artifacts. An example of this problem
was the apparent discovery of a decreasing period in {\it RXTE} data from
IRAS~18325$-$5926 that was eventually demonstrated to be due to
detector background (Fabian et al.\ 1998a,b).  A final problem that
this author has personal experience with is that of source confusion.
{\it EXOSAT} and {\it Ginga} data revealed a 12,000 second periodicity
in the Seyfert 1 galaxy NGC 6814 (e.g., Leighly et al.\ 1994).  Imaging
observations later showed that the signal was dominated by a
cataclysmic variable star only 40 arcminutes away (Madejski et al.\
1993). 

These misadventures do not, however, preclude the possibility that
periodicity or quasiperiodicity could be present in AGN light curves.
It may be that we simply are not looking in the right place, or that
we don't have the right data.  In this contribution, I discuss the
significant (marginal) detection of periodicity X-ray data from three
(four) Seyfert 1 galaxies.

\section{Methodology}  

A standard Monte Carlo procedure was used to detect periodicity in the
AGN light curves, with small differences between the procedure used in
Halpern, Leighly \& Marshall (2003), which discusses analysis and
results of the {\it EUVE} light curves, and in Moran et al.\ 2005,
which discusses the {\it Chandra} result.  First, the slope and
normalization of the assumed underlying power-law continuum was
determined.  In Halpern, Leighly \& Marshall (2003), 100 long light
curves were generated, using the method of Timmer \& K\"onig (1995),
for each point on a grid of trial power-law slopes and normalizations.
These long light curves were resampled and rebinned in the same way as
the real light curves, appropriate Poisson noise was added, and the
periodogram was computed and rebinned.  The average and standard
deviation periodogram from the situations at each point in the grid
was then compared with that of the real data using $\chi^{2}$.  The
minimum $\chi^2$ was assumed to locate the best-fitting values of the
slope and normalization of the continuum.  The procedure used for the
analysis of the {\it Chandra} data from NGC 4395 differed in one
respect. Light curves were produced for a range of slopes, but only
one normalization.  Following Vaughan, Fabian \& Nandra (2003), the
result was scaled to different normalizations, and appropriate Poisson
noise computed and added.

Knowing the shape of the power spectrum continuum, 10,000 suitably
rebinned and sampled light curves were generated using the method of
Timmer \& K\"onig (1995).  The power spectra from these were used to
assessed the local confidence level of the detection, first by
counting the number of incidents in which the power of the simulations
exceeded the power of the real data, and second by comparing with a
$\chi^2_2$ distribution (e.g., Leahy et al.\ 1983).  As discussed in
Benlloch et al.\ (2001), one must account for the number of independent
frequencies searched to obtain an estimate of the global significance
by multiplying the single-trial confidence by that number.  

\section{Results and Discussion}  

Halpern, Leighly \& Marshall (2003) present an atlas of {\it EUVE}
light curves from Seyfert galaxies.  Three objects had exceptionally
long light curves: 33 days for Ton S180, 26 days for 1H~0419$-$577,
and 20 days for RX~J0437.1$-$4711.  Periodicity analysis as outlined
above revealed evidence for a 2.08-day signal in Ton S180 with global
confidence level of 98\%, a 0.908-day signal in RX~J0437.4$-$4711, and
a 5.8-day signal in 1H~0419$-$577 with a global confidence level of
only 64\%.  Interestingly, the periodic signal in RX~J0437.1$-$4711
appears to be transient; periodicity analysis of the first and second
halves of the light curve show no detectable signal, and a signal at
0.89-day with global confidence of 96\%, respectively.

Moran et al.\ (2005) present analysis of the {\it Chandra} data from
the low-luminosity Seyfert 1 galaxy NGC 4395.  A period of 396 seconds
was found in the light curve with estimated global significance of
97.6\%.  In this case, we again found evidence that the signal is
transient; no signal appears in the first half of the observation, but
in the second half a strong signal appears with estimated global
confidence of 99.95\% at a period of 396 seconds.

\begin{figure} 
\centerline{\includegraphics[width=3.5truein]{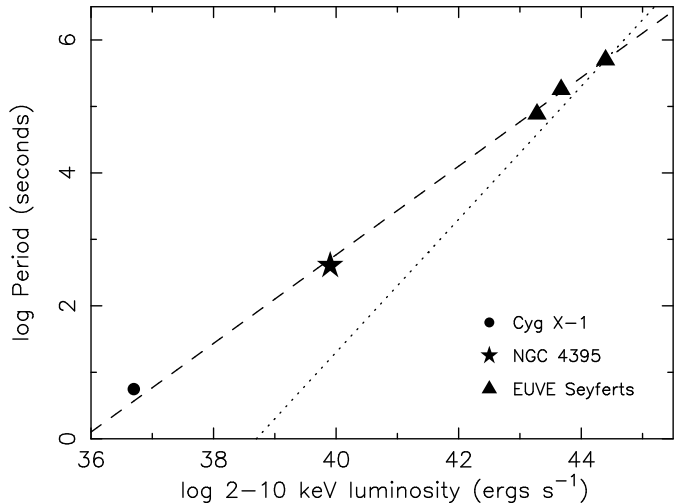}}
\caption[]{Log of the period as a function of the log of the 2--10 keV
  luminosity for 4 AGN and Cyg X-1.  The dotted and dashed lines have
  slopes of 1 and 2/3, respectively.}
\end{figure}

It is interesting to note that the presence of a QPO does not appear
to depend on Seyfert type.  Ton~S180 is a narrow-line Seyfert 1 galaxy,
1H~0419$-$577 is a Seyfert 1.5, and both NGC 4395 and
RX~J0437.1$-$4711 are both classified as ordinary Seyfert 1 galaxies.  

As noted in Halpern, Leighly, \& Marshall (2003), there is a
correspondence between the X-ray luminosity and the period such that  
more luminous objects have longer periods.  This correspondence is
expected if the luminosity and characteristic time scales  both scale
with the black hole mass.  Things get interesting when I compare the
period and luminosity of NGC~4395 and Cyg~X-1 in the hard state, as
shown in Fig.\ 1.   For Cyg~X-1, we use the pre-{\it RXTE} period of 5.6~s
(e.g., Tanaka 1995), but note that using the {\it RXTE} values of
0.5--1~s (e.g., Gilfanov et al.\ 2000) would not change this figure
much because it is logarithmic.  Remarkably, the linear
relationship between the log of the period and luminosity is
maintained over the 8 decades of luminosity between Cyg~X-1 and
1H~0419$-$577.  However, the slope of the  relationship is $2/3$,
rather than 1, as would be expected for fixed $L/L_{Edd}$ and emission
$R/R_S$.  One possibility is that the QPO is associated with the
truncation radius of the optically thick, geometrically thin accretion
disk, and the dependence on luminosity is a secondary effect, in that
the Seyferts are radiating closer to $L/L_{Edd}$ than does Cyg~X-1 in
the hard state, in which $L/L_{Edd}=0.02$.  If that were the case, and
if $R=3R_S$ for Ton~S180, we infer that $R=23R_S$ for NGC~4395, and
$R=95R_S$ for Cyg~X-1.  Interestingly, $R\approx 100R_G$ is
approximately the inferred radius of the inner edge of the optically
thick accretion disk when Cyg~X-1 is in the hard state (e.g.,
Gilfanov, Churazov \& Revnivtev 2000).   

\begin{figure} 
\centerline{\includegraphics[width=4.5truein]{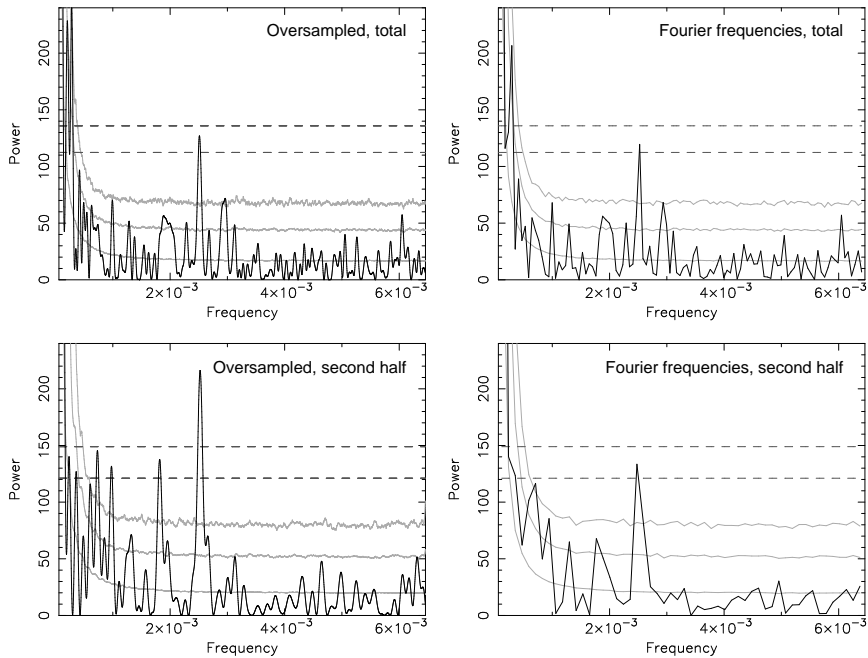}}
\caption[]{Power spectra for the NGC 4395 {\it Chandra} observation
  (Moran et al.\ 2005).  In each plot, the lower three light grey
  lines show the 68, 95, and 99\% single-trial significance levels
  from the simulated data, while the upper two dark grey dashed
  lines show the 95 and 99\% global significance levels computed
  assuming Poisson noise continuum.}
\end{figure}

\section{Caveats and Cautions}  

What do these results say about the incidence of periodicity in X-ray
light curves from AGN? There are thousands of X-ray light curves that
have not been tested for periodicity, so one might argue that these
are isolated incidents and not representative.  However, as discussed
in Halpern, Leighly, \& Marshall (2003), if the characteristic period
of Seyfert 1 galaxies is one day or longer, most of those thousands of
light curves will be not suitable for periodicity searches.  The {\it
EUVE} light curves were exceptionally long, and thus the fact that
some evidence for periodicity was found in the three longest ones
suggests perhaps that periodicity is common in AGN.  Then what about
NGC~4395?  It is notable for being the lowest luminosity Seyfert 1,
with a 2--10 keV luminosity of only $8 \times 10^{39}\rm \,
erg\,s^{-1}$ (Moran et al.\ 2005), much lower than a typical Seyfert 1
galaxy.  The fact that we observe the period to scale with the
luminosity in the {\it EUVE} sample suggests that NGC~4395 should have
an exceptionally low period; therefore, since it is unique,
luminosity-wise, it makes sense that it is unique, period-wise.

On the other hand, we observe that the periodicity appears to be
transient in both NGC 4395 and RX~J0437.1$-$4711 in that in both cases
it is detected only in the second halves of the light curves.  This
suggests that  even if periodicity were common in AGN, it may be
difficult to observe even in much better data than currently available,
simply because too few cycles will be present to detect.

It has been suggested by Vaughan (2005) that the global confidence of
the detections claimed in Halpern, Leighly \& Marshall (2003) and in
Moran et al.\ (2005) have been overestimated for two reasons.  The
first reason is that we oversample the periodogram; this was done to
improve the sensitivity to frequencies other than the usual Fourier
frequencies.  This is potentially a problem because oversampling the
periodogram will produce more higher peaks, and thus comparison of the
real data, with spurious high peaks, with the simulated data will
result in overly high single-trial probabilities, or false detections.
We investigate this problem in Fig.\ 2, for the NGC~4395 data. We find
almost the same values for the global significance for the total light
curve, possibly because the peak for the oversampled power spectrum is
only 20\% from one of the Fourier frequencies.  For the second half of
the light curve, where the periodicity appears stronger, the
significance of the oversampled peak is much higher than for the
Fourier peak.  Perhaps this is because that peak lies almost exactly
in between Fourier frequencies.

The second criticism raised by Vaughan (2005) is that the standard
analysis does not take into account the uncertainty in the shape of the
underlying continuum.  That is, we assume that once we measure the
underlying power-law spectral continuum, it is fixed, whereas in
reality there is uncertainty in the slope, normalization, and even the
Poisson noise level.  This criticism appears to be well founded, and
will no doubt lower the global confidence level of the periodicity
detections. Quantifying the new global confidence levels is somewhat
difficult, though; there is no direct way to do this.

\acknowledgements
KML gratefully acknowledges helpful discussions with Jules
Halpern and Andrzej Zdziarski, thanks the organizers for the
opportunity to speak on this topic, and acknowledges funding from NASA
ADP grant NAG5-13110.

\end{article}

\begin{thebibliography}{}


\bibitem[\protect\citeauthoryear{Benlloch et al.}{2001}]{benllochetal}
Benlloch, S., J.~Wilms, R.~Edelson, T.~Yaqoob, and R.~Staubert.
\newblock{Quasi-periodic Oscillation in Seyfert Galaxies: Significance Levels. The Case of Markarian 766}.
\newblock{ApJ}, 562, 121, 2001.

\bibitem[\protect\citeauthoryear{Boller et al.}{2001}]{bolleretal}
Boller, Th., R.~Keil, J.~Tr\"umper, J., P.~T.~O'Brien, J.~Reeves, and M.~Page.
\newblock{Detection of an X-ray periodicity in the Narrow-line Seyfert 1 Galaxy Mrk 766 with XMM-Newton}.
\newblock{A\&A}, 365, L146, 2001.

\bibitem[\protect\citeauthoryear{Fabian et al.}{1998}]{fabianetala}
Fabian, A.\ C., J.~C.~Lee, K.~Iwasawa, K.~Jahoda, W.~N.~Brandt, and C.~S.~Reynolds.  
\newblock{IRAS 18325$-$5926}.
\newblock{IAUC}, 6871, 3, 1998b.

\bibitem[\protect\citeauthoryear{Fabian et al.}{1998}]{fabianetalb}
Fabian, A.\ C., J.~C.~Lee, K.~Iwasawa, W.~N.~Brandt, and C.~S.~Reynolds.  
\newblock{IRAS 18325$-$5926}.
\newblock{IAUC}, 6835, 3, 1998a.

\bibitem[\protect\citeauthoryear{Gilfanov, Churazov \& Revnivtev}{2000}]{Gilfanovetal}
Gilfanov, M., E.~Churazov, and M.~Revnivtsev.
\newblock{Frequency-resolved spectroscopy of Cyg X-1: fast variability
of the reflected emission in the soft state}.
\newblock{MNRAS}, 316, 923, 2000.

\bibitem[\protect\citeauthoryear{Halpernetal}{2003}]{halpern}
Halpern, J.~P., K.~M.~Leighly, and H.~L.~Marshall.
\newblock{An Extreme Ultraviolet Explorer Atlas of Seyfert Galaxy
  Light Curves: Search for Periodicity}.
\newblock{ApJ}, 585, 665, 2003.

\bibitem[\protect\citeauthoryear{Leahy et al.}{1983}]{leahyetal}
Leahy, D.~A., W.~Darbro, R.~F.~Elsner, M.~C.~Weisskopf, S.~Kahn,
P.~G.~Sutherland, and J.~Grindlay.
\newblock{On searches for pulsed emission with application to four
  globular cluster X-ray sources -- NGC 1851, 6441, 6624, and 6712}.
\newblock{ApJ}, 266, 160, 1983.

\bibitem[\protect\citeauthoryear{Leighly et al.}{1994}]{leighlyetal}
Leighly, K., H.~Kunieda, Y.~Tsusaka, H.~Awaki, and S.~Tsuruta.
\newblock{Evidence for X-ray flux and spectral modulation by absorption in NGC 6814. 1: The nature of the most rapid variability}.
\newblock{ApJ}, 421, 69, 1994.

\bibitem[\protect\citeauthoryear{Madejski et al.}{1993}]{madeskietal}
Madejski, G., C.~Done, T.~J.~Turner, R.~F.~Mushotzky, P.~Serlemitsos,
F~Fiore, M.~Sikora, and M.~Begelman.
\newblock{Solving the Mystery of the Periodicity in the Seyfert Galaxy NGC 6814}.
\newblock{Nature}, 365, 626, 1993.

\bibitem[\protect\citeauthoryear{Moran et al.}{2005}]{moranetal}
Moran, E.~C., Eracleous, M., Leighly, K.~M.~Leighly, G.~Chartas,
A.~V.~Filippenko, L.~C.~Ho, and P.~R.~Blanco.
\newblock{Extreme X-ray Behavior and Quasi-Periodic Oscillations in
  the Low-Luminosity Active Nucleus of NGC 4395}.
\newblock{ApJ}, submitted, 2005.

\bibitem[\protect\citeauthoryear{Tanaka}{1995}]{tanaka}
Tanaka, Y.
\newblock{Physics of Neutron Stars and Black Holes}.
\newblock{In {\em X-ray Binaries},
  Cambridge University Press}, 1995.

\bibitem[\protect\citeauthoryear{Timmer \& Konig}{1995}]{timmer}
Timmer, J., and M.~K\"onig
\newblock{On Generating Power Law Noise}.
\newblock{A\&A}, 300, 707, 1995.

\bibitem[\protect\citeauthoryear{Vaughan}{2005}]{vaughan}
Vaughan, S.
\newblock{A simple test for periodic signals in red noise}.
\newblock{A\&A}, in press, 2005.

\bibitem[\protect\citeauthoryear{Vaughan, Fabian, Nandra}{2003}]{vaughanetal}
Vaughan, S., A.~C.~Fabian, and K.~Nandra.
\newblock{X-ray continuum variability of MCG--6-30-15}.
\newblock{MNRAS}, 339, 1237, 2003.

\end{thebibliography}
\end{document}